\documentstyle[preprint,tighten,aps]{revtex}

\begin{document}
\draft
\preprint{IFUP-TH 65/97}
\title{
The three-loop $\beta$-function of SU(N) lattice gauge
theories with Wilson fermions. }
\author{C. Christou$^a$, A. Feo$^b$, H. Panagopoulos$^a$, E. Vicari$^b$}
\address{
$^a$Department of Natural Sciences, University of Cyprus,
P.O.Box 537, Nicosia CY-1678, Cyprus.}
\address{
$^b$Dipartimento di Fisica dell'Universit\`a 
and I.N.F.N., 
Piazza Torricelli 2, I-56126 Pisa, Italy.}


\maketitle

\begin{abstract}
We calculate the third coefficient of the lattice
$\beta$-function associated with the Wilson formulation
for both gauge fields and fermions.
This allows us to evaluate the three-loop correction 
(linear in $g_0^2$) to the 
relation between the lattice $\Lambda$-parameter
and the bare coupling $g_0$,
 which is important in order to verify
asymptotic scaling predictions.
Our calculation also leads to the two-loop relation 
between the coupling renormalized
in the $\overline{{\rm MS}}$ scheme and $g_0$.

The original version of this paper contained a numerical error in 
one of the diagrams, which has now been corrected. The calculations,
as well as the layout of the paper have remained identical, but there
are some important changes in the numerical results.


\end{abstract}

\newpage


\section{Introduction}
\label{introduction}

In lattice formulations of asymptotically free field theories, such as QCD,
the relevant region 
for the continuum physics is that where scaling is verified.
Scaling in the continuum limit requires that
all RG-invariant dimensionless ratios of physical quantities 
approach their continuum value with
non-universal corrections which get depressed
by integer powers of the inverse correlation length.
This can be equally stated by introducing a lattice scale
$\Lambda_L$, which is a particular solution of the RG equation 
\begin{equation}
\left( -a{\partial\over \partial a} + \beta_L(g_0)
{\partial\over \partial g_0} \right)\Lambda_L = 0
\end{equation}
($a$ is the lattice spacing), i.e. 
\begin{equation}
a \Lambda_L = \exp\left[ - \int^{g_0} {dg\over \beta_L(g)}\right],
\end{equation}
and requiring that the ratio of any RG invariant
quantity to the appropriate power of $\Lambda_L$ approach
 a constant in the continuum limit $g_0\rightarrow 0$.
$\beta_L(g_0)$ is the lattice $\beta$-function 
\begin{equation}
\beta_L(g_0)= -a{dg_0\over da} \mid_{\rm physical \;\; quantities}, 
\label{lattb}
\end{equation}
which tells us how the bare coupling and the lattice
spacing must be changed in order to keep physical quantities fixed.

In the asymptotic region 
$g_0\rightarrow 0$ one may perform a weak coupling
expansion of the $\beta$-function
\begin{equation}
\beta_L(g_0) =
-b_0 g^3_0 -b_1 g_0^5 - b_2^{L}g_0^7 + ...,
\end{equation}
where, in $SU(N)$ gauge theory with $N_f$ fermion species,
\begin{eqnarray}
&&b_0 = {1\over (4\pi)^2} 
\left({11\over 3}N-{2\over 3}N_f\right),\\
&&b_1= {1\over (4\pi)^4} \left[{34\over 3}N^2 - N_f \left(
{13\over 3}N- {1\over N}\right)\right].
\end{eqnarray}
Consequently one has the asymptotic relation
\begin{equation}
a\Lambda_L = \exp \left( -{1\over 2b_0g_0^2}\right)
(b_0g_0^2)^{-{b_1/2b_0^2}}
\left[ 1 + q g_0^2 + O\left(g_0^4\right)\right],
\label{asympre}
\end{equation}
where 
\begin{equation}
q = {b_1^2-b_0b_2^L\over 2b_0^3}.
\label{qq}
\end{equation}

The verification of the two-loop behavior of $\Lambda_L$
is usually called asymptotic scaling. 
Knowledge of the first correction to the two-loop 
approximation of $\Lambda_L$ is important in order to verify
the asymptotic prediction  (\ref{asympre}).
We recall that Monte Carlo simulations using the Wilson action are
actually performed at $g_0\simeq 1$, and so deviations
from the two-loop formula might not be negligible.
$b_2^L$ has been recently calculated for the pure gauge 
theory~\cite{LWpaper,A-F-P}.
In this paper we extend this calculation to the full theory
including fermions in the Wilson formulation.

The knowledge of $b_2^L$ can be also used
to improve the pertubative relation between the coupling $g$
renormalized
in the $\overline{{\rm MS}}$ 
(modified minimal subtraction renormalization) scheme 
and the bare lattice coupling
$g_0$, which is
 useful in calculations such as studies
concerning running couplings
(see e.g. Refs.~\cite{L-etal-cou1,L-etal-cou2,deDetal,AHPPPR}).
Actually, as we shall see, in order to calculate $b_2^L$, 
we will first compute the two-loop relation between 
$g$ and $g_0$. This, supplemented
with knowledge of the three-loop coefficient of the 
$\beta$-function in the $\overline{{\rm MS}}$  
scheme, allows us to derive $b_2^L$.

In Sec.~\ref{sec2} we describe the method we employed to calculate
$b_2^L$. We present the results of our lattice 
perturbative calculations, from which 
we derive the three-loop
coefficient $b_2^L$ associated with  the
Wilson formulation of both gauge fields and fermions.
Sec.~\ref{sec3} describes in some detail 
the necessary lattice perturbative calculations we performed
in the background 
gauge framework.

\section{Perturbative calculation of the $\protect\bbox{\beta}$-function.}
\label{sec2}

The lattice $\beta$-function depends on the lattice formulation considered.
We recall that only the first two coefficients of its perturbative 
expansion are universal,
i.e. $b_0$ and $b_1$. 
We consider the Wilson formulation
for both the pure gauge theory and fermions:
\begin{eqnarray}
&&S_L =  {1\over g_0^2} \sum_{x,\mu,\nu}
{\rm Tr}\left[ 1 - U_{\mu\nu}(x) \right]  +
\sum_{f}\sum_{x} (4r+m_{f,0})\bar{\psi}_{f}(x)\psi_f(x) + 
\nonumber \\
&& -{1\over 2}\sum_{f}\sum_{x,\mu}
\left[ 
\bar{\psi}_{f}(x)\left( r - \gamma_\mu\right)
U_{\mu}(x)\psi_f(x+\hat{\mu})+
\bar{\psi}_f(x+\hat{\mu})\left( r + \gamma_\mu\right)
U_{\mu}(x)^\dagger
\psi_{f}(x)\right] ,
\label{latact}
\end{eqnarray}
where $U_{\mu\nu}(x)$ is the usual product of link variables
$U_{\mu}(x)$ along the perimeter of a plaquette
originating at $x$ in the positive $\mu$-$\nu$ directions;
$r$ is the Wilson parameter; $f$ is a flavor index.  

The lattice $\beta$-function is independent of the
renormalized fermionic masses.
Indeed, since it is dimensionless,
its mass dependence may come only through
the combination $x\equiv am$, where $m$ is the renormalized
fermionic mass.
Since lattice gauge theories with Wilson fermions are 
renormalizable~\cite{Reisz},
the lattice $\beta$-function must have a well defined 
zero lattice-spacing limit. Thus any dependence on $x$,
and therefore on $m$, must disappear
in that limit. In the presence of fermions with different masses,
one may also assume a well defined chiral limit in order
to complete this argument. 
So in order to calculate $\beta_L(g_0)$ we can restrict 
ourselves to the case of $N_f$ fermions with
zero renormalized mass.

The relation between the bare lattice coupling $g_0$ 
and the renormalized coupling $g$ defined in a 
generic renormalization scheme is
\begin{equation}
g_0 = Z_g(g_0,a\mu) g,
\label{grenorm}
\end{equation}
where $\mu$ indicates a renormalization scale.
$Z_g(g_0,a\mu)$ depends on the renormalization scheme.
The lattice $\beta$-function (\ref{lattb}) can be derived from $Z_g$
by
\begin{equation}
\beta^L(g_0) = -g_0 a {d\over da} \ln Z_g(g_0,a\mu)\mid_{\mu,g}.
\label{lbf}
\end{equation}
Thus the evaluation of $b_2^L$ would require
the calculation of $Z_g(g_0,a\mu)$ up to three loops
on the lattice.

The lattice calculation can be simplified by 
exploiting the following considerations.
Let us choose
the $\overline{\rm MS}$ procedure as renormalization scheme
in which the renormalized coupling $g$ is defined.
The corresponding $\beta$-function 
\begin{equation}
\beta(g) = \mu{dg\over d\mu} \mid_{\rm bare\;\; quantities} 
=-b_0 g^3 -b_1 g^5 - b_2 g^7 + ...
\label{cbf}
\end{equation}
is known to three loops. 
By writing the $\overline{\rm MS}$ $\beta$-function in the form
\begin{equation}
\beta(g)=
-g \mu{d\over d\mu} \ln Z_g(g_0,a\mu)\mid_{a,g_0},
\label{cbf2}
\end{equation}
and by comparing it with Eq.~(\ref{lbf}),
one can easily derive the relation
\begin{equation}
\beta^{L}(g_0) = 
\left( 1 - g_0^2 {\partial\over \partial g_0^2} \ln Z_g^2 \right)^{-1}
Z_g \beta(g_0Z_g^{-1}),
\label{lbcb}
\end{equation}
which is valid to all orders of perturbation theory.
We write $Z_g^2$ as
\begin{equation}
Z_g(g_0,a\mu)^2  = 1 + L_0(a\mu) g_0^2 + 
L_1(a\mu) g_0^4 + O(g_0^6)
\label{zgzg}
\end{equation}
where $L_0(x) = 2b_0 \ln x + l_0$ and 
$L_1(x) = 2b_1 \ln x + l_1$.
The constant $l_0$ is related to the ratio 
of the $\Lambda$ parameters 
associated with the particular
lattice regularization and the
$\overline{\rm MS}$ renormalization scheme:
\begin{equation}
l_{0} = 2b_0\ln \left( \Lambda_L/
\Lambda_{\overline{\rm MS}}\right).
\end{equation}
Its value as a function of $N$ and $N_f$ can be inferred
from Refs.~\cite{Hasenfratz,Kawai}. It has been obtained
 with improved accuracy 
in Ref.~\cite{LWpaper} (pure gluons) and in the present paper
(fermionic case). For $r=1$
\begin{equation}
l_0 ={1\over 8N} - 0.16995599 N + 0.00669600 N_f.
\label{l0}
\end{equation}
By expanding Eq.~(\ref{lbcb}) in powers of $g_0^2$,
one finds the well-known result that 
the coefficients $b_0$ and $b_1$ are the same in $\beta^L(g_0)$ and 
$\beta(g)$. 
Furthermore 
\begin{equation}
b_2^L= b_2 -b_1l_{0}+ b_0 l_{1}.
\label{b2lrel}
\end{equation}
Thus, since $b_2$ is known~\cite{Tarasovetal}, 
\begin{equation}
b_2 = {1\over (4\pi)^6}
\left[ {2857\over 54}N^3 + N_f \left( -{1709N^2\over 54} + {187\over 36}
+{1\over 4N^2}\right) + 
N_f^2 \left( {56 N \over 27} - {11\over 18N}\right)\right],
\label{b2MS}
\end{equation}
the evaluation of $b_2^{L}$ requires only
a two-loop calculation on the
lattice, i.e. the calculation of the constant $l_{1}$.

The computation of
the $\overline{\rm MS}$ renormalization constant 
$Z_g$ is easier in the background field gauge.
In fact, this renormalization constant
has a simple relationship with the
background field renormalization constant
$Z_A$~\cite{Abbott},
\begin{equation}
Z_A(g_0,a\mu)   Z_g(g_0, a\mu)^2 = 1.
\label{eq:zeqza}
\end{equation}
L\"uscher and Weisz~\cite{LWbg}
have shown that pure lattice gauge theory
with a background gauge field is renormalizable to all 
orders in perturbation theory. No additional counterterms
are required besides those already needed in the absence of 
a background field. Their argument, based on 
renormalizability of pure lattice gauge theory,
BRS, background gauge and shift symmetries of the 
lattice functional integral,
can be extended to full lattice QCD in the Wilson
formulation. An essential point is the renormalizability of lattice
gauge theory with Wilson fermions proved by Reisz to all orders
in perturbation theory~\cite{Reisz}.

As a consequence of the relation (\ref{eq:zeqza}),
in order to calculate $Z_g$
we need to compute only the two-loop self-energy of the background
field on the lattice.
In the background field formulation~\cite{ellism}  the
links are written as
\begin{eqnarray}
U_{\mu}(x) &=& U^{q}_{\mu}(x) U^{cl}_{\mu}(x), \nonumber \\
U^{q}_{\mu}(x) &\equiv& e^{i g_0 Q_{\mu}(x)}, \nonumber \\
U^{cl}_{\mu}(x) &\equiv& e^{i a g_0 A_{\mu}(x)},
\end{eqnarray}
where $Q_{\mu}(x)=T^c Q^c_{\mu}(x)$ and
$A_{\mu}(x) = T^c A_{\mu}^c(x)$ are the quantum and
background fields respectively.
The following gauge-fixing term preserves gauge invariance of the
background field
\begin{eqnarray}
S_{gf} &=& \lambda_0 \sum_{\mu , \nu} \sum_{x}
\hbox{Tr} \, D^-_{\mu} Q_{\mu}(x) D^-_{\nu} Q_{\nu}(x), \\
D^-_{\mu} Q_{\nu}(x) &\equiv & U^{cl-1}_{\mu}(x -
{\hat \mu}) Q_{\nu}(x - {\hat \mu}) U^{cl}_{\mu}(x - {\hat \mu}) -
Q_{\nu}(x).
\end{eqnarray}
We worked in the Feynman gauge, $\lambda_0 = 1$.
This gauge fixing produces the following
Fadeev Popov action for the ghosts fields $\omega$ and $\overline\omega$
\begin{eqnarray}
S_{gh} &=& 2 \sum_{x} \sum_{\mu} \hbox{Tr} \,
(D^+_{\mu}\omega(x))^{\dagger} \Bigl( D^+_{\mu}\omega(x) +
i g_0 \left[Q_{\mu}(x),
\omega(x)\right] + \case{1}{2}
i g_0 \left[Q_{\mu}(x), D^+_{\mu}\omega(x) \right] \nonumber\\
 & & \quad - \case{1}{12}
g_0^2 \left[Q_{\mu}(x), \left[ Q_{\mu}(x),
D^+_{\mu}\omega(x)\right]\right] + \cdots \Bigr), \\
D^+_{\mu}\omega(x) &\equiv & U^{cl}_{\mu}(x) \omega(x + {\hat \mu})
U^{cl-1}_{\mu}(x) - \omega(x).
\end{eqnarray}
Finally the change of integration variables from links to vector
fields yields a jacobian that can be rewritten as 
a new term $S_m$ in the action:
\begin{equation}
S_{m} = \frac{1}{12} N g_0^2 \sum_{x} \sum_{\mu} \hbox{Tr} \,
Q_{\mu}(x) Q_{\mu}(x) + \cdots
\end{equation}
In all the expansions above we have written only
the relevant terms for our two-loop computation.

The full action is therefore 
\begin{equation}
S = S_L + S_{gf} + S_{gh} + S_m.
\end{equation}
The vertices needed in perturbation theory are obtained as usual
by expanding the exponential of the action $S$ in powers of
the fields.

As in Ref.~\cite{LWpaper},
we set the following notation.
We rewrite the renormalized one-particle irreducible
two-point functions of the background and quantum fields as
\begin{eqnarray}
\Gamma^{AA}_R(p)^{ab}_{\mu\nu} &=& 
-\delta^{ab}\left( \delta_{\mu\nu}p^2 - p_\mu p_\nu\right)
\left( 1 - \nu_R(p)\right)/g^2, \\
\Gamma^{QQ}_R(p)^{ab}_{\mu\nu} &=& 
-\delta^{ab}
\left[ \left( \delta_{\mu\nu}p^2 - p_\mu p_\nu\right)
\left( 1 - \omega_R(p)\right) + \lambda p_\mu p_\nu \right], \\
\nu_R(p) &=& \sum_{l=1}^\infty g^{2l} \nu_R^{(l)}(p) ,\\
\omega_R(p) &=& \sum_{l=1}^\infty g^{2l} \omega_R^{(l)}(p). 
\end{eqnarray}
Correspondingly on the lattice 
\begin{eqnarray}
\sum_\mu \Gamma^{AA}_L(p)^{ab}_{\mu\mu} &=& 
-\delta^{ab}3\widehat{p}^2 
\left[ 1 - \nu(p)\right]/g_0^2, \\
\sum_\mu \Gamma^{QQ}_L(p)^{ab}_{\mu\mu} &=& 
-\delta^{ab}\widehat{p}^2 
\left[ 3\left( 1 - \omega(p)\right) + \lambda_0\right], \\
\nu(p) &=& \sum_{l=1}^\infty g_0^{2l} \nu^{(l)}(p) ,\\
\omega(p) &=& \sum_{l=1}^\infty g_0^{2l} \omega^{(l)}(p). 
\end{eqnarray}

The bare and renormalized functions
are related by
\begin{equation}
\left[ 1 - \nu_R(p,\mu,g) \right]
= Z_A \left[ 1 - \nu(p,a,g_0) \right],
\label{eq:selfs}
\end{equation}
therefore
\begin{equation}
Z_g^2 = {1 - \nu(p,a,g_0)\over 1 - \nu_R(p,\mu,g)}.
\label{zgzgzg}
\end{equation} 
In order to compare continuum against lattice expressions,
one also needs to renormalize 
the gauge parameter 
according to $\lambda=Z_Q\lambda_0$~\cite{Abbott},
where $Z_Q$ is the renormalization constant of the quantum
field. $Z_Q$ is needed only to
one loop and can be evaluated by imposing
\begin{equation}
\left[ 1 - \omega_R(p,\mu,g) \right] = Z_Q 
\left[ 1 - \omega(p,a,g_0) \right].
\end{equation}
One can fix $\lambda_0=1$, but then 
one must take $\lambda=Z_Q$ as corresponding
renormalized gauge parameter.

The $\overline{\rm MS}$ renormalized functions 
necessary for the calculation of $Z_g$ to two loops are
\begin{eqnarray}
\nu_R^{(1)}(p,\lambda)=&&
{N\over 16\pi^2} \left[  
-{11\over 3}\ln {p^2\over \mu^2} +{205\over 36}
+{3\over 2\lambda} + {1\over 4 \lambda^2}\right] +
{N_f\over 16\pi^2}\left[ {2\over 3}\ln{p^2\over \mu^2}
- {10\over 9}\right],
\label{nu1}\\
\omega_R^{(1)}(p,\lambda)=&&
{N\over 16\pi^2} \left[ \left(-{13\over 6}+{1\over 2\lambda}\right) 
\ln{p^2\over \mu^2} +{97\over 36}
+{1\over 2\lambda} + {1\over 4 \lambda^2}\right] +
{N_f\over 16\pi^2}\left[ {2\over 3}\ln{p^2\over \mu^2}- 
{10\over 9}\right],
\label{omega1}\\
\nu_R^{(2)}(p,\lambda=1)=&&
{N^2\over \left(16\pi^2\right)^2} 
\left[ -8\ln{p^2\over \mu^2} +{577\over 18}-6\zeta(3)\right]+\nonumber \\
&&{N_f\over \left(16\pi^2\right)^2}
\left[ N\left( 3 \ln{p^2\over \mu^2} - {401\over 36}\right)+ {1\over N}
\left(-\ln{p^2\over \mu^2} + {55\over 12} - 4\zeta(3)\right)\right].
\label{nu2b}
\end{eqnarray}

Details of the lattice calculations will be given in the next section.
Here we just list the results. We report only 
the fermionic contributions for $r=1$
which must be added
to the pure gauge functions calculated in Refs.~\cite{LWpaper,A-F-P}.
\begin{eqnarray}
\nu^{(1)}(p,\lambda_0=1) = &&\nu^{(1)}(p,\lambda_0=1)\mid_{N_f=0} + 
N_f \left[{1\over 24\pi^2}\ln (a^2p^2) + k_{1f} \right],
\\
\omega^{(1)}(p,\lambda_0=1) =&& \omega^{(1)}(p,\lambda_0=1)\mid_{N_f=0} + 
N_f \left[{1\over 24\pi^2}\ln (a^2p^2) + k_{1f} \right],\\
\nu^{(2)}(p,\lambda_0=1) = &&\nu^{(2)}(p,\lambda_0=1)\mid_{N_f=0} 
+ N_f \left[
{1\over (16\pi^2)^2} \left( 3N - {1\over N}\right)
\ln (a^2p^2) + k_{2f}{1\over N} + k_{3f} N\right].
\end{eqnarray}
For $r=1$ we found
\begin{eqnarray}
k_{1f} =&-& 0.013732194(5),\nonumber \\
k_{2f} =& & 0.0011877(14),\nonumber \\
k_{3f} =&-& 0.0013617(16) .
\label{results}
\end{eqnarray}
The origin and meaning of the errors in the above results
will be explained in the next section.

We have now what we need to calculate $Z_g$ to two loops
using Eq.~(\ref{zgzgzg}). The resulting two-loop constant $l_1$,
cf. Eq.~(\ref{zgzg}), is given by\footnote{
Here and below, results pertaining to the pure gluonic
case can be obtained with higher precision from Ref.~\cite{LWpaper}.}
\begin{eqnarray}
l_1 =&& -{3\over 128 N^2} + 0.018127763 - 0.0079101185 N^2 \nonumber \\
&&+ N_f\left[ -0.0011967(14){1\over N} + 0.0009998(16) N\right].
\end{eqnarray}
Through Eq.~(\ref{b2lrel}) one can then obtain $b_2^L$.
For example for $N=3$ we find 
\begin{equation}
b_2^L=-0.0015998323 + 0.0000799(4) N_f - 0.00000605(2) N_f^2.
\end{equation}

Knowledge of $b_2^L$ allows us
to evaluate the coefficient $q$ of the linear
correction to asymptotic scaling, cf. Eq.~(\ref{qq}).
For $N=3$, we obtain:
\begin{equation}
q\simeq 0.1896 \quad (N_f=0), \qquad 
q\simeq 0.2160 \quad (N_f=2), \qquad q\simeq 0.2355 \quad (N_f=3).
\end{equation}
These numbers show that for $g_0\simeq 1$, which is a typical value of
the bare coupling where simulations are nowadays performed,
the linear correction to asymptotic scaling cannot be ignored.

Finally, the relation between 
the $\overline{\rm MS}$ coupling $\alpha\equiv g^2/(4\pi)$
and $\alpha_0\equiv g_0^2/(4\pi)$
can be easily read from Eq.~(\ref{zgzg}):
\begin{equation}
\alpha = \alpha_0 + d_1(a\mu)\alpha_0^2 +
d_2(a\mu)\alpha_0^3+O\left( \alpha_0^4\right),
\end{equation}
where, for $r=1$,
\begin{eqnarray}
d_1(x)=&& -4\pi L_0(x) \nonumber\\
=&&-{1\over 2\pi} 
\left({11\over 3}N-{2\over 3}N_f\right)\ln x
-{\pi\over 2N} + 2.13573007 N - 0.08414443(8) N_f,
\label{d1}
\end{eqnarray}
and
\begin{eqnarray}
d_2(x) =&& (4\pi)^2\left[ L_0(x)^2-L_1(x)\right]\nonumber\\
=&& d_1(x)^2 -
{1\over 24 \pi^2} \left[ 34 N^2 - N_f \left(
13 N- {3\over N}\right)\right]\ln x  \nonumber \\
&&+  
{3\pi^2\over 8N^2} - 2.8626216 + 1.2491158 N^2 + 
N_f \left[ 0.18898(22) {1\over N} - 0.15789(26) N\right].
\end{eqnarray}

For comparison we have also estimated $b_2^L$ for 
$r=0.5$ and $r=2.0\,$. For $N=3$, we obtain
\begin{equation}
b_2^L-b_2^L|_{N_f=0}
\simeq + 0.000092 N_f - 0.0000033 N_f^2
\end{equation}
for $r=0.5$, and
\begin{equation}
b_2^L-b_2^L|_{N_f=0}
\simeq + 0.0000079 N_f - 0.0000039 N_f^2
\end{equation}
for $r=2.0\,$.

\section{The Calculation in Lattice Perturbation Theory}
\label{sec3}

\noindent
{\it i) Preliminaries}
\smallskip

In this Section we proceed to describe the technical aspects in our
calculation of the quantities $\nu^{(1)}(p)$, $\nu^{(2)}(p)$. The
fermionic contributions in $\omega^{(1)}(p)$ are identical to those in
$\nu^{(1)}(p)$.

Two diagrams containing fermions contribute to $\nu^{(1)}(p)$, shown
in Figure 1. There are 18 two-loop fermionic diagrams contributing to
$\nu^{(2)}(p)$, as well as two diagrams containing an insertion of the
one-loop fermion mass counterterm; these are shown in Figure 2.

The algebra involving lattice quantities was performed using a
symbolic manipulation package in Mathematica, developed by us in
recent years. For the purposes of the present work, this package was
extended to include fermions.

A first, relatively brief, part in the evaluation of the diagrams is
the contraction. This is done completely automatically for diagrams
with an arbitrary number of loops, once the types of vertices, as well
as the ``incidence matrix'' for the diagram are specified. This step
also includes: Complete reduction of color structures, Dirac matrices
and tensor structures; exploiting permutation symmetry and lattice
rotational invariance to keep the size of the expression down to a
minimum; use of trigonometry to arrive at a canonical form. The
resulting expression is a rational function of sines and cosines of
the external ($p$) and internal ($q, k$) momenta.

\medskip
\noindent
{\it ii) Extracting the external momentum}
\smallskip

The next task is to make explicit the functional dependence of each
diagram on $p$. The two-loop amplitude $\nu^{(2)}(p)$ can be written as:
\begin{equation}
\nu^{(2)}(p) = \nu^{(2)}(p)|_{N_f{=}0} + \sum_i \nu_i(p)
\end{equation}
(the index $i$ runs over diagrams with fermions shown
in Fig.2), where, generally,
\begin{equation}
\widehat{ap}^2 \nu_i(p) = c_{0,i} + c_{1,i}\, a^2 \sum_\mu {p_\mu^4 \over
p^2} + a^2 p^2 \left\{ c_{2,i} \left({\ln a^2 p^2 \over (4 \pi)^2}\right)^2 +
c_{3,i}\, {\ln a^2 p^2 \over (4 \pi)^2} + c_{4,i} \right \} + 
O((ap)^4) 
\end{equation}
($\widehat{p}^2 = 4 \sum_\mu \sin^2(p_\mu/2)$). The dependence of
$c_{n,i}$ of $N, N_f$ is:
\begin{equation}
c_{n,i} = \left[ c_{n,i}^{(-1)} /N + c_{n,i}^{(1)} N\right] N_f
\end{equation}
We note in passing that certain diagrams are infrared convergent only
when taken in pairs: (7, 11), (8, 18), (9, 17); we have evaluated
these accordingly, with due care taken to avoid divergences in intermediate
results.

To extract the $p$-dependence, we first isolate the superficially
divergent terms; these are responsible for the double
logarithms. There are relatively few such terms, and in the pure
gluonic case they all have been tabulated~\cite{A-F-P}. We can use
these tables also in diagrams with fermions, applying successive
subtractions of the type: 
\begin{equation}
{1\over \overline{q}^2} = 
{1\over \widehat{q}^2} + \left( {1\over \overline{q}^2} -
{1\over \widehat{q}^2}\right)
\end{equation}
where $\overline{q}^2$ is the 
inverse fermionic propagator: $\overline{q}^2 =
(\widehat{q}^2\, r/2)^2 + \sum_\mu \sin^2 q_\mu\,$. This leads to the
tabulated expressions plus a series of superficially convergent terms.

The rest of the expression may still contain subdivergences, which can
give rise to single logarithms.  These are handled using successive
subtractions of the type:
\begin{equation}
{1\over {\widehat{q+k}}^2} = {1\over {\,\widehat{k}}^2} + \left({1\over
{\widehat{q+k}}^2} - {1\over {\,\widehat{k}}^2}\right)
\end{equation}
The first term on the right hand side then yields factorized
one-loop expressions, whose $p$-dependence is easily extracted, while
the second term is more convergent. Finally, all remaining
terms, containing no (sub-)divergences can be evaluated by Taylor
expansion in $ap$. 

In our code, judicious choices for the right set of subtractions are
applied automatically to each term. While the extraction of
divergences is not particularly complicated conceptually, it leads to
a great proliferation in the resulting number of terms. As an example,
diagram 16 by itself leads to ${\sim} 100$ {\it types} of expressions;
each type must then be numerically evaluated on its own (because it
contains a different set of subtracted propagators) and contains
typically up to some hundreds of terms.

\medskip
\noindent
{\it iii) Numerical integration}
\smallskip

At this stage, all the above types of expressions no longer contain
$p$ and must be numerically integrated over the internal momenta. The
integration is done in momentum space over finite lattices;
an extrapolation to infinite size is then performed (see below).

For expressions with only gluonic propagators a coordinate space
method was proposed in~\cite{coord}, yielding high precision
results. Recently, this method was also used for some study cases
involving fermions~\cite{coordf}. For the present calculation,
involving fermionic and bosonic propagators with multiple
subtractions, the coordinate space method becomes a bit more
complicated; we may return to it in the future, if the need arises for
more precision than presented in the present paper.

Fortran code for the numerical evaluation of each type of expression
is then created automatically by our programs. The code is highly
optimized: It avoids redundant evaluation of common subexpressions,
of symmetric regions in momentum space, of $r$-independent
subexpressions, etc. The code can be run for different values of $r$
and lattice size $L\,$. For the present paper we used $L\le 30$;
we present $r=1$ and, for comparison, we also estimated
$r=0.5$, $r=2.0\,$.

Extrapolation to infinite lattice size is of course a source of
systematic error. To estimate this error, our procedure
carries out automatically the following steps: First, different
extrapolations are performed using a broad spectrum of functional
forms of the type:
\begin{equation}
\sum_{i,j} e_{i,j}\, L^{-i}\, (\ln L)^j
\end{equation}
For the $k^{\rm th}$ such extrapolation, a deviation $d_k$ is
calculated using several criteria for quality of fit. Finally, these
deviations are used to assign weights $d_k^{-2}/(\sum_k d_k^{-2})$ to
each extrapolation, producing a final value together with
the error estimate. We have checked the validity of these estimates in
cases where the exact answer was known (see also below), finding the
estimate to be always correct.

\medskip
\noindent
{\it iv) Cancellations and cross checks}
\smallskip

Several constraints exist on the coefficients $c_{n,i}$. We have used
these constraints as verifications both on the algebraic expressions
and on the numerical results:

\smallskip
\noindent
{\underline {$c_{0,i}$}}: Gauge invariance requires 
\begin{equation}
\sum_i c_{0,i} = 0.
\label{c0i}
\end{equation} 
We checked this property in three ways: Firstly, a formal,
Ward-identity type derivation was performed, in which vertices with
background fields at zero momentum were written in terms of
appropriate derivatives of inverse propagators. From this we find not
only Eq.(\ref{c0i}), but also some additional constraints:
\begin{eqnarray}
&&  2\, c_{0,1} + c_{0,3} = 0 \nonumber \\
&& c_{0,1} + c_{0,3} + c_{0,6} + c_{0,8} = 0 \nonumber \\
&& (c_{0,4} + c_{0,5})|_{N^2 = 2} = 0 \nonumber \\
&& c_{0,7} + c_{0,9} + c_{0,11} + \case{1}{2}c_{0,16}
+ c_{0,17} = 0 \nonumber \\
&& c_{0,13} = 0 \nonumber \\
&& c_{0,19} + c_{0,20} = 0 \nonumber \\
&& (c_{0,12} + c_{0,18}) (N^2 -2) + c_{0,14} (N^2-1) - c_{0,15}
(N^2-1)(N^2-2) = 0
\label{c0other}
\end{eqnarray}
The above are true for any value of the fermion mass $m$ and Wilson
parameter $r$. Secondly, all the above identities were checked
by algebraic manipulation of the full expression for each of the
coefficients. Thirdly, we substituted in each identity the numerical
results for the coefficients, finding in all cases zero within the
error estimates.

\smallskip
\noindent
{\underline {$c_{1,i}$}}: The sum of these terms must vanish if Lorentz
invariance is to be recovered in the continuum limit. Only diagrams 9
and 17 give nonvanishing contributions, and we checked in the same
ways as above that their sum is zero.

\smallskip
\noindent
{\underline {$c_{2,i}$}}: These coefficients must coincide with those
of the continuum. We checked that this is so:
\begin{equation} 
c_{2,15} = {1\over 3}{1\over N} N_f\, , \qquad
c_{2,16} = {4\over 3} N N_f\, , \qquad
c_{2,17} = -{5\over 3} N N_f\, , \qquad
c_{2,18} = {1\over 3} {N^2-1\over N} N_f\, . 
\end{equation}
For all other diagrams: $c_{2,i} = 0$.

\smallskip
\noindent
{\underline {$c_{3,i}$}}: The total contribution for single logarithms
must coincide with the continuum result:
\begin{equation}
\sum_i c_{3,i} = {1\over 16 \pi^2} (3 N - {1\over N}) N_f
\end{equation}
Again, this was checked both algebraically and numerically.

\medskip
The results for each diagram are presented in Tables 1 and 2. Diagrams
not appearing in these Tables give vanishing contributions.\footnote{
Compared to the original version, only one entry is different,
corresponding to diagram 19.}

For the one-loop amplitude $\widehat{ap}^2\,\nu^{(1)}(p)$, the individual
contributions from the corresponding two diagrams in Figure 1 are,
respectively: 
\begin{equation}
0.040848920(4) - a^2 p^2\, 0.013732194(5) + {1\over 24 \pi^2} \, a^2
p^2\, \ln(a^2 p^2)\, ,  \qquad  -0.040848919(5)
\end{equation}

\medskip
In concluding this section, we would like to point out that the
procedure outlined here applies unchanged to several other interesting
cases, including: More complicated actions, matrix elements of
different operators. We hope to address these issues in a future
publication.

\bigskip\bigskip\bigskip
\noindent
{\bf Acknowledgements:} H. P. would like to acknowledge the warm
hospitality extended to him by the Theory Group in Pisa during various
stages of this work.


\begin{table}
\caption{Coefficients $c^{(-1)}_{0,i}$, $c^{(-1)}_{3,i}$,
$c^{(-1)}_{4,i}\,$. $r=1$. 
\label{tab1}}
\begin{tabular}{cr@{}lr@{}lr@{}l}
\multicolumn{1}{c}{$i$}&
\multicolumn{2}{c}{$c^{(-1)}_{0,i}$}&
\multicolumn{2}{c}{$c^{(-1)}_{3,i}$}&
\multicolumn{2}{c}{$c^{(-1)}_{4,i}$}\\
\tableline \hline
1 &$-$0&.00158221542(13)&0& & 0&\\
3 &0&.0031644309(6)&0&.051644463410 &$-$0&.0010637877(8) \\
4 &0&.00039273(5)&0& &0& \\
5 &$-$0&.0005077(9)&0& &0&.00010879(5) \\
6 &$-$0&.0086173230(12)&0& &0& \\
$7+11$&0& &0& &0&.00069292(12) \\
8 &0&.0070351064(17)&0&.258222317052 &$-$0&.0031633707(22) \\
12 &0&.00069739(6)&0& &0& \\
14 &0&.00022988(14)&0&.00528244566(5) &0&.0000757(12) \\
15 &$-$0&.00060983(5) &$-$0&.019554587(16) &0&.0001732(6) \\
18 &$-$0&.0002019(4) &0&.023786891(14) &$-$0&.00007910(15) \\
19 &$-$0&.010721166(2)&$-$0&.325714118(9) &0&.004443346(3) \\
20 &0&.0107211662(6)&0& &0& \\
\end{tabular}
\end{table}

\begin{table}
\caption{Coefficients $c^{(1)}_{0,i}$, $c^{(1)}_{3,i}$, 
$c^{(1)}_{4,i}\, $. $r=1$.
\label{tab2}}
\begin{tabular}{cr@{}lr@{}lr@{}l}
\multicolumn{1}{c}{$i$}&
\multicolumn{2}{c}{$c^{(1)}_{0,i}$}&
\multicolumn{2}{c}{$c^{(1)}_{3,i}$}&
\multicolumn{2}{c}{$c^{(1)}_{4,i}$}\\
\tableline \hline
1 &0&.00158221542(13) &0& &0& \\
3 &$-$0&.0031644309(6)&$-$0&.051644463410&0&.0010637877(8) \\
4 &$-$0&.00039273(5) &0& &0& \\
5 &0&.0004503(6) &0& &$-$0&.00008893(4) \\
6 &0&.0086173230(12) &0& &0& \\
$7+11$&0&.0014560(8) &0& &$-$0&.0007427(4) \\
8 &$-$0&.0070351064(17)&$-$0&.258222317052 &0&.0031633707(22) \\
$9+17$&$-$0&.0020085(5)&0&.08132606(7) &$-$0&.0011335(13) \\
12 &$-$0&.00069739(6) &0& &0& \\
13 &0& &0& &0&.00006908(5) \\
14 &$-$0&.00011494(7) &$-$0&.00264122283(3) &$-$0&.0000378(6) \\
16 &0&.0011052(4) &$-$0&.0517470(17) &0&.0007092(7) \\
18 &0&.0002019(4) &$-$0&.023786891(14) &0&.00007910(15) \\
19 &0&.010721166(2) &0&.325714118(9) &$-$0&.004443346(3) \\
20 &$-$0&.0107211662(6) &0& &0& \\
\end{tabular}
\end{table}

\begin{figure}
\caption{
Diagrams with fermionic lines contributing
to the one-loop function $\nu^{(1)}(p)$.
Dashed lines ending on a cross represent background gluons.
Solid lines represent fermions.
}
\label{fig1}
\end{figure}

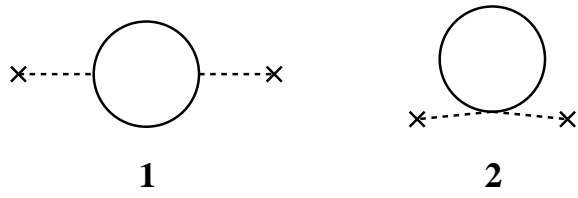
\begin{figure}
\caption{
Diagrams with fermionic lines contributing
to the two-loop function $\nu^{(2)}(p)$.
Dashed lines represent gluonic fields;
those ending on a cross stand for background gluons.
Solid lines represent fermions.
The filled circle is a one-loop
fermion mass counterterm. 
}
\label{fig2}
\end{figure}

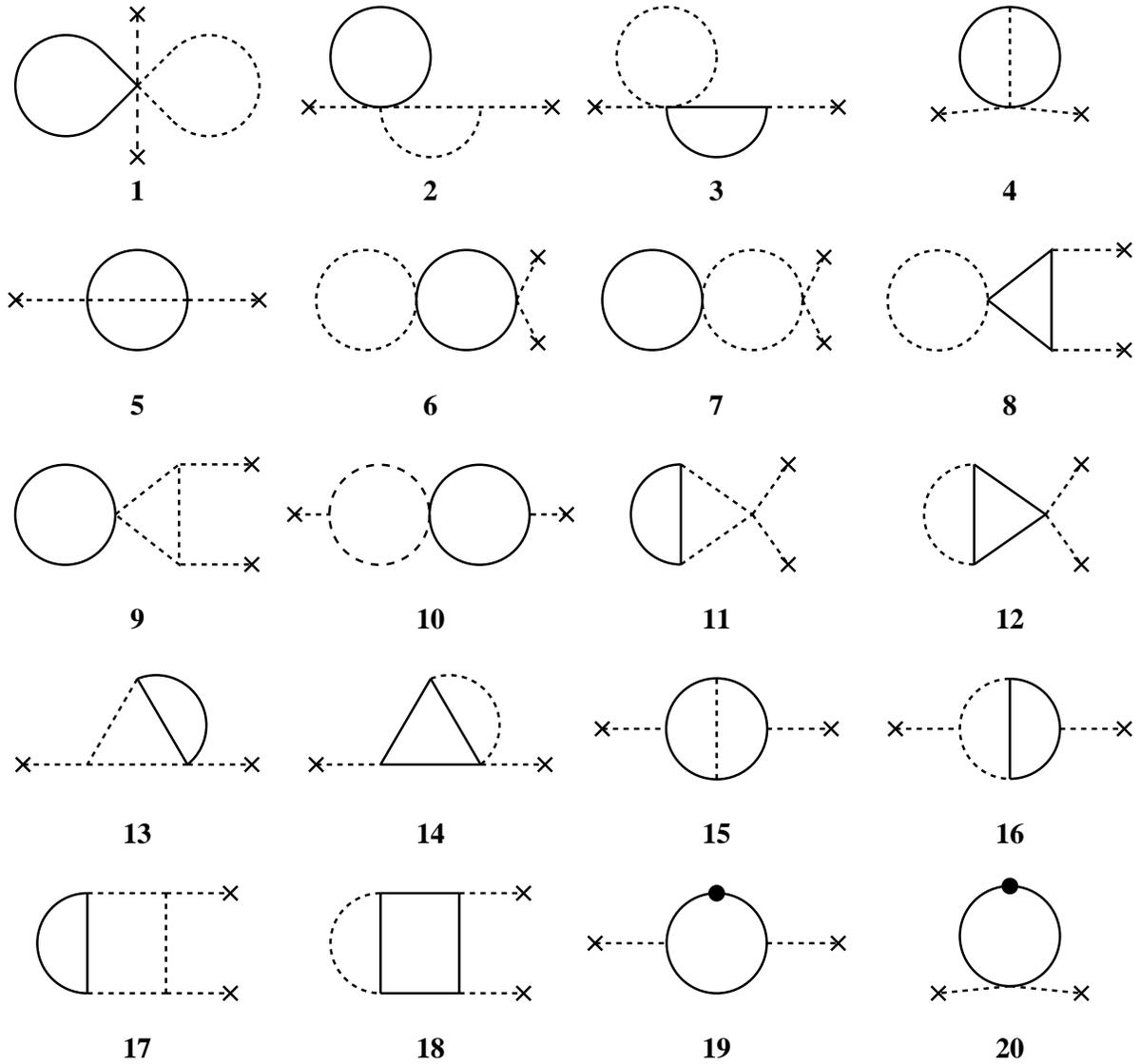
\end{document}